# The effect of the glass transition in fullerite $C_{60}$ on Ar impurity diffusion

A.V. Dolbin, V.B. Esel'son, V.G. Gavrilko, V.G. Manzhelii, N.A.Vinnikov, and R.M. Basnukaeva

B. Verkin Institute for Low Temperature Physics and Engineering of the National Academy of Sciences of Ukraine

47 Lenin Ave., Kharkov 61103, Ukraine

E-mail: dolbin@ilt.kharkov.ua

**Abstract**.

The kinetics of sorption and subsequent desorption of argon gas by a $C_{60}$ powder has been investigated in the temperature interval 58-290 K. The temperature dependence of the coefficients of Ar diffusion in fullerite has been obtained using the measured characteristic times of sorption. The diffusion coefficients of Ar decrease monotonically with lowering temperature in the whole range of the investigated temperatures, which corresponds to the thermally activated diffusion of Ar atoms in fullerite. The glass transition in fullerite induces an order-of-magnitude decrease in the activation energy of Ar diffusion in fullerite. Most likely this is because new directions may appear due to the glass transition in which the barriers separating the interstitial voids in the $C_{60}$ lattice are significantly lower.

## Introduction

At room temperature $C_{60}$ molecules form a molecular crystal-fullerite which has a face-centered cubic (fcc) lattice and the Fm3m symmetry [1]. The noncentral interaction of $C_{60}$ molecules is much weaker than the central one. As a result, in the solid phase the $C_{60}$ molecules can change their orientations at quite low temperatures. As room temperature the rotation of $C_{60}$ molecules at the lattice sites is weakly retarded. As the temperature decreases to $T_c \approx 260$ K, a structural orientational phase transition occurs and the symmetry of the FCC lattice of fullerite lowers from Fm3m to Pa3. On a further drop of temperature to $T_g \approx 90$ K the rotation of the $C_{60}$ molecules about the <111> axes is almost fully frozen, short of complete orientational ordering. This is how an orientational glass having no long-range orientational order of $C_{60}$ molecules is formed.

In the fullerite lattice each $C_{60}$ molecule has two tetrahedral and one octahedral interstitial voids, their average linear sizes being ~2.2 Å and ~ 4.2 Å, respectively [2]. The occupancy of the voids by the impurity particles in the process of impurity diffusion in $C_{60}$ has a significant effect on the properties of fullerite, especially when it is in the state of orientational glass [3,4]. It is known that smaller tetrahedral voids are accessible for sorption of He atoms [5,6] and octahedral voids are open for larger atoms and molecules. It was found that cooling $C_{60}$ below $T_g$ essentially affected the temperature dependences of the coefficients of $^4$He, $^3$He, $H_2$ and Ne diffusion in fullerite. The diffusion coefficients of these substances decreased with lowering temperature [6-8]. In the case of $^4$He and $^3$He the diffusion coefficients were independent of temperature below 8 K. This behavior of the diffusion coefficients suggests that the rate of sorption of light impurities by fullerite $C_{60}$ is determined by the process of tunneling of impurity particles in $C_{60}$, which is dominant at low temperatures. The sharp changes in the temperature dependences of the diffusion coefficients observed for all of the investigated impurity particles (helium isotopes, hydrogen, neon) occurred near the temperature of the glass transition. The reason may be that in the course of the orientational glass transition some directions appear in the $C_{60}$ lattice along which the barriers between the interstitial voids hindering the motion of impurity molecules are lower.

Thus, the transition of fullerite from the phase with partial orientational ordering to the state of orientational glass enhances the probability of tunneling and its contribution to the diffusion processes, In turn the lowering of the barriers should relax the temperature dependence of the classical thermally activated contribution to impurity diffusion. To test this assumption, we investigated the diffusion of a heavier impurity for which the effect of quantum phenomena on the kinetics of $C_{60}$ saturation is negligible. At the same time the saturation vapor pressure of such impurity at $T<T_g$ is sufficient to apply the experimental technique used in [6-8]. In this study the doping element was Ar. The sorption – desorption kinetics of Ar atoms by a $C_{60}$ powder was investigated in the temperature interval 58-290 K. The kinetics of $C_{60}$ saturation with Ar was investigated at room temperature by the X-ray diffraction method [9]. No investigations of the kinetics of Ar sorption by $C_{60}$ at lower temperatures have been reported so far.

## Experimental technique

The kinetics of sorption-desorption of Ar gas was investigated by measuring the time dependence of the pressure of the gas contacting a $C_{60}$ powder in a closed volume. The details of the technique and the experimental facility are described elsewhere [6,7]. The used $C_{60}$ powder had grains of ~ 1 µm, 99,99 wt. % purity and a mass of 618.21 mg. Directly before the investigation the powder was evacuated for 72 hours at $T \sim 450$ °C to remove possible gas impurities and moisture. Then during a short period of time (~ 30 min) it was transferred in the air to the measuring cell and evacuated again at room temperature for 48 hours. The $C_{60}$ sample was saturated with chemically pure Ar gas (> 0.005 % $O_2$).

The kinetics of Ar sorption-desorption in fullerite was investigated in the temperature interval 58-290 K. The saturation of $C_{60}$ with Ar gas was performed at $P_{Ar}$=10 Torr and $T \geq 65$ K. The Ar desorption from the $C_{60}$ sample was performed at the lowest temperature of the experiment after removing quickly (30 sec) the Ar gas from the measuring cell and its cjjling to 58 K. In the all saturation-desorption runs the pre-assigned temperature of the sample was maintained constant and the pressure of Ar gas in the measuring cell was considerably below the saturation vapor pressure of Ar for a particular pre-assigned temperature of the sample. These saturation conditions permitted us to avoid condensation of the Ar vapor grains and the cell walls.

The pressure variations of the Ar gas inside the closed volume of the cell with the cell $C_{60}$ sample during saturation and desorption were registed using a capacitive pressure sensor (MKS «Baratron»). The measurement error was no more than 0.05%. When the sorption at each pre-assigned sample temperature was completed , the remainder of Ar gas was removed quicly (~1 min) from the measuring cell. Then the cell was hermetically sealed again and the pressure variations of the subsequent Ar desorption from $C_{60}$ were measured. After desorption the sample temperature was brought to the next pre-assigned value and the saturation- desorption procedure was repeated.

## Results and discussion

The obtained time dependences of the Ar pressure variations in the cell with a $C_{60}$ sample in the process of sorption or desorption are well described by the exponential one-parameter function of Eq. (1) (see Fig.1). This suggests that the adsorbed Ar atoms fill mainly the octahedral subsystem of voids in $C_{60}$, which is consistent with the X-ray structural data in [10,11]:

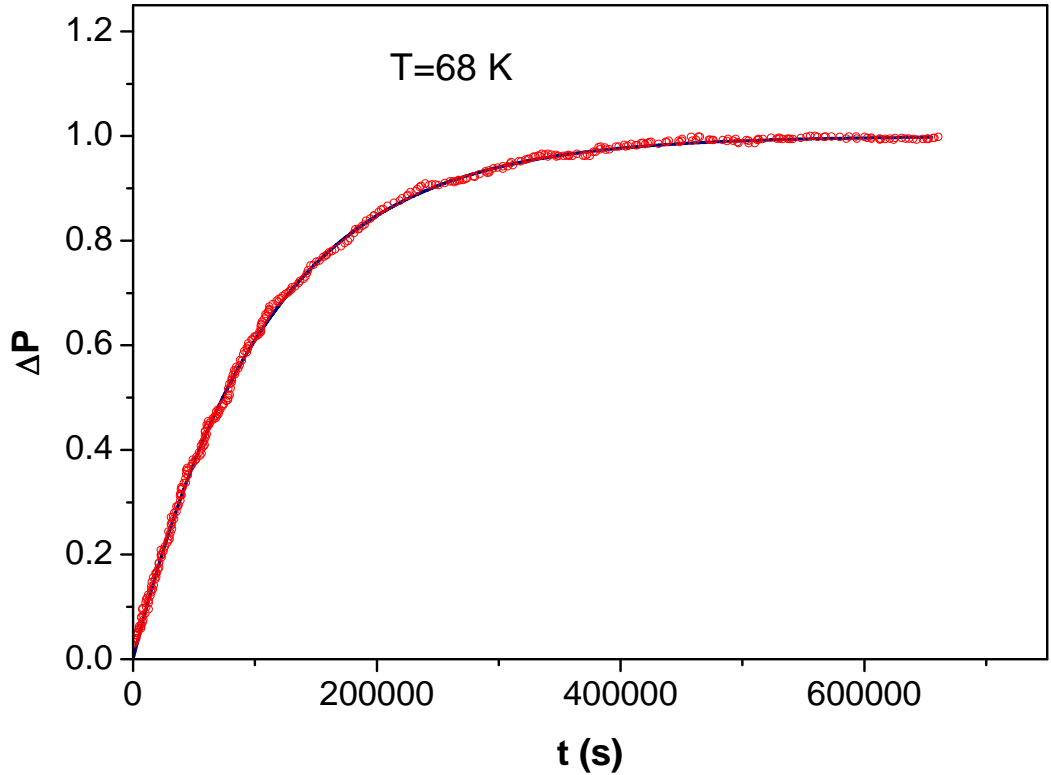

Fig.1. The pressure variations in the course of Ar desorption from a $C_{60}$ sample (symbols, experimental data) and their description by Eg.(1)(Line). As an example, the results obtained at the temperature of the sample $T = 60$ K are included.

$$\Delta P = A \cdot (1 - \exp(-t/\tau)) \qquad (1)$$

The exponent $\tau$ and the parameter A were found by fitting to the experimental results. It may be assumed that the exponent $\tau$ corresponds to the characteristic time during which the Ar atoms occupy the octahedral voids in the $C_{60}$ lattice. Note that the characteristic times of Ar sorption and desorption taken at each pre-assigned temperature of the sample coincided within the measurement error. Besides, the $\tau$ – values were practically independent of the starting Ar pressure on $C_{60}$ saturation in our experiment. The resulting Ar concentration in the saturated sample indicated that up to 15% of the octahedral voids were occupied in $C_{60}$. According to our estimates, this quantity of Ar exceeds the sorptive capacity of the grain surfaces in a $C_{60}$ crystal by overran order of magnitude (the typical grain size in our sample was ~1 μm). It is likely that the adsorbed Ar atoms occupy mainly the inside of the $C_{60}$ grains and the sorption – desorption kinetics is determined by Ar diffusion in the $C_{60}$ crystal. Using the characteristic sorption – desorption times and Eq. (2), we obtained the coefficients $D$ of the Ar diffusion in fullerite $C_{60}$ and plotted their temperature dependences (see Fig.2, solid triangles).

$$D \approx \frac{\overline{\ell}^2}{6\tau} \qquad (2),$$

where $\overline{\ell}$ is the average grain size in the $C_{60}$ powder (~1μm); $\tau$ is the characteristic time of sorption.

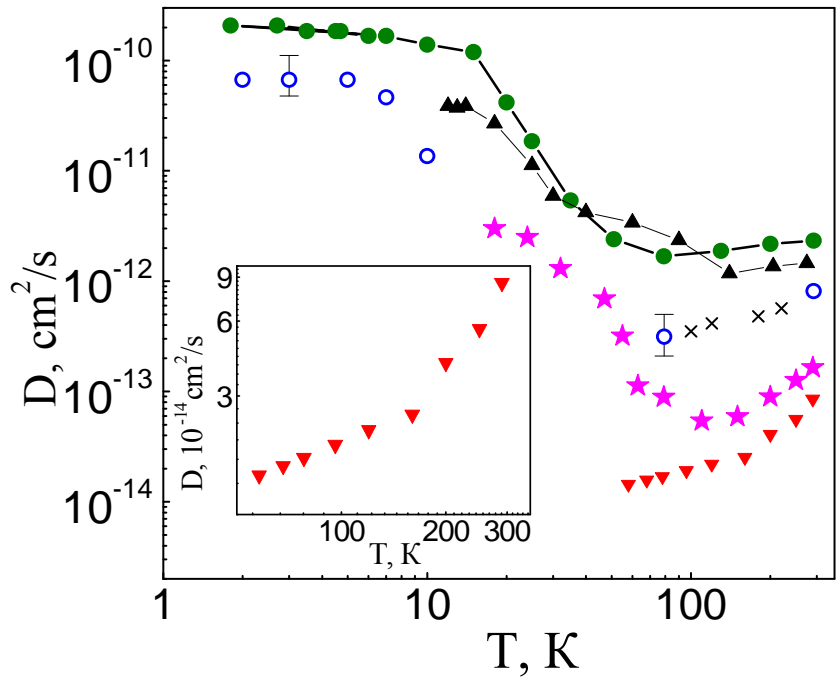

Fig.2. The temperature dependence of the coefficients of Ar diffusion in fullerite $C_{60}$ (solid triangles). For comparison, the figure carries the temperature dependences of the diffusion coefficients of $^4$He (empty circles [6], crosses, this study), $^3$He (solid circles [7]), $H_2$ (empty triangles [8]) and Ne (asterisks [8]) for the octachedral subsystems of voids in fullerite.

Since we measured the pressure of the sorbed gas, the relatively low tension of the saturated vapor of Ar prohibited an adequate-accuracy investigation of the saturation-desaturation kinetics below 58 K.

In contrast to the previously investigated impurities, the diffusion coefficients of Ar decrease with lowering temperature in whole T-interval of the experiments, which corresponds to the process of thermally activated Ar diffusion in $C_{60}$ (see Fig.2). This behavior of the diffusion coefficients may be attributed to the larger mass and effective diameter of Ar atoms in comparison with $H_2$ He and Ne, which suggests rather low probability of Ar tunneling through the void-separating barriers in the $C_{60}$ lattice.

To find the activation energy of diffusion $E_a$, the temperature dependence of the coefficients of the Ar diffusion in fullerite $C_{60}$ was plotted in the coordinates $Y=ln(D)$ vs. $X=1/T$

(Fig.3). The dependence $Y(X)$ is linear if the diffusion process corresponds to the Arrhenius equation:

$$D = D_0 \exp\left(-\frac{E_a}{k_B T}\right) \quad (3),$$

where $E_a$ is the activation energy of diffusion, $D_0$ is the entropy factor dependent on the frequency of collisions of the host and impurity molecules; $k_B$ is the Boltzmann constant.

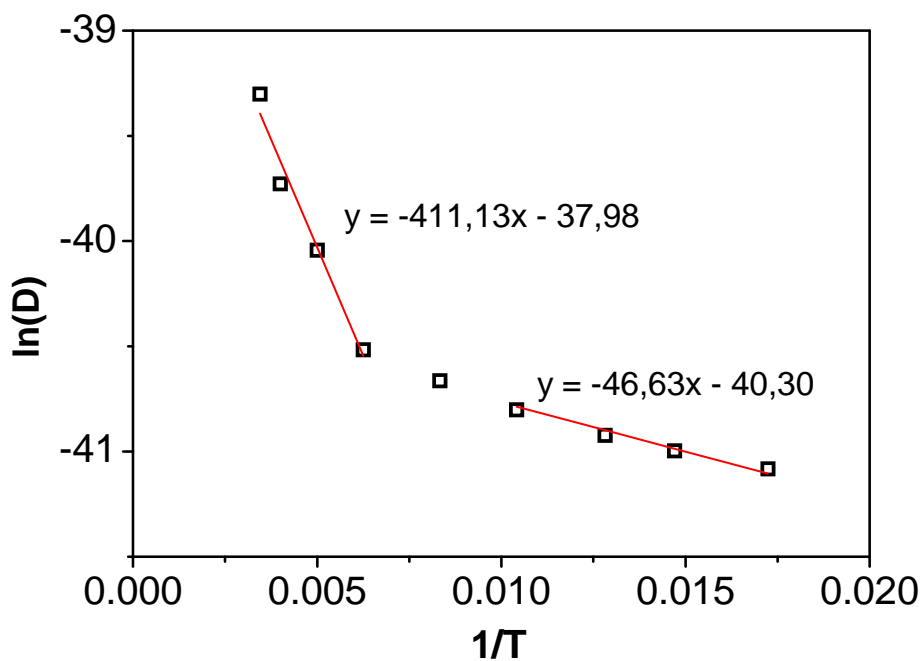

Fig.3. The linear portions of dependence $Y=Ln(D)$ vs. $X=1/T$ of the coefficients of Ar diffusion in fullerite $C_{60}$. The region of the variation of the angular coefficients of the line corresponds approximately to the temperature interval of the glass transition in fullerite.

It is seen in Fig. 3 that the dependence $Y(X)$ has two distinct linear portions. The region, where the angle of the line slope changes corresponds approximately to the temperature interval of the glass transition in fullerite.

Thus, the glass transition in fullerite causes a sharp (an order of magnitude) decrease in the activation energy of the Ar diffusion in fullerite $C_{60}$ (see Fig. 4a). The conclusion supports the assumption [7] that the glass transition can produce some directions in which the barriers separating the interstitial voids in the $C_{60}$ lattice are significantly lower. This enhances the probability of tunneling and its contribution to the diffusion processes of light impurities (He, $H_2$, Ne). Simultaneously, the glass transition relaxes the temperature dependence of thermally activated diffusion, which shows up as a bend in the temperature dependence of the diffusion coefficients of the Ar impurity near $T_g$ (see Figs. 2, 3).

We estimated the activation energies of diffusion for the He, $H_2$ and Ne impurities in fullerite using the temperature dependences of the diffusion coefficients obtained in previous investigations [6 – 8] (see Fig. 4) in the interval $T = 100 – 290$ K. Equation (3) fails at lower temperatures because of increasing contribution of quantum effects to diffusion. To obtain reliable estimates of the activation energy for the $^4$He impurity, we carried out additional investigations of its saturation-desorption kinetics in fullerite in interval $T = 100 – 220$ K (Fig. 2, crosses).

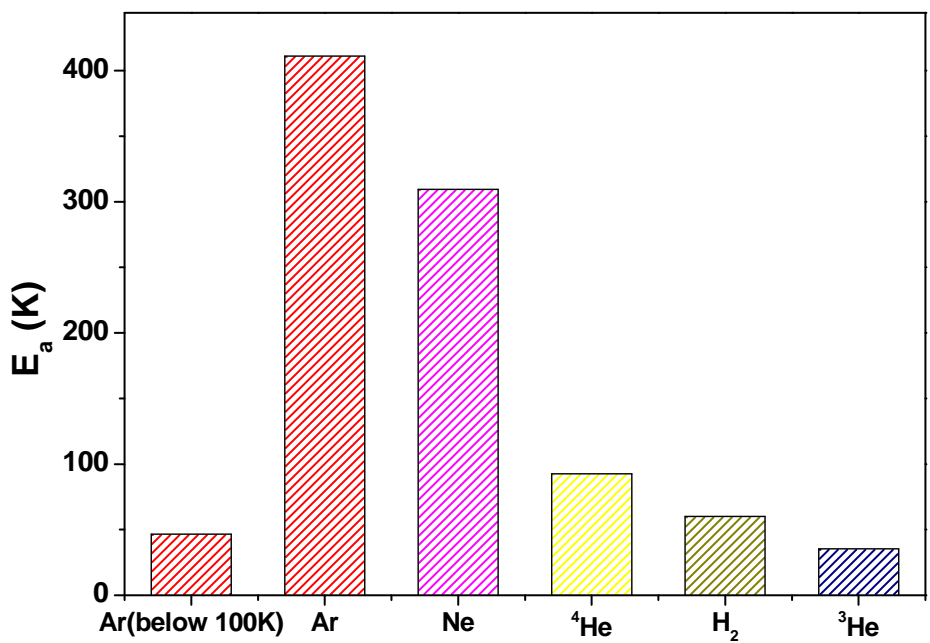

a)

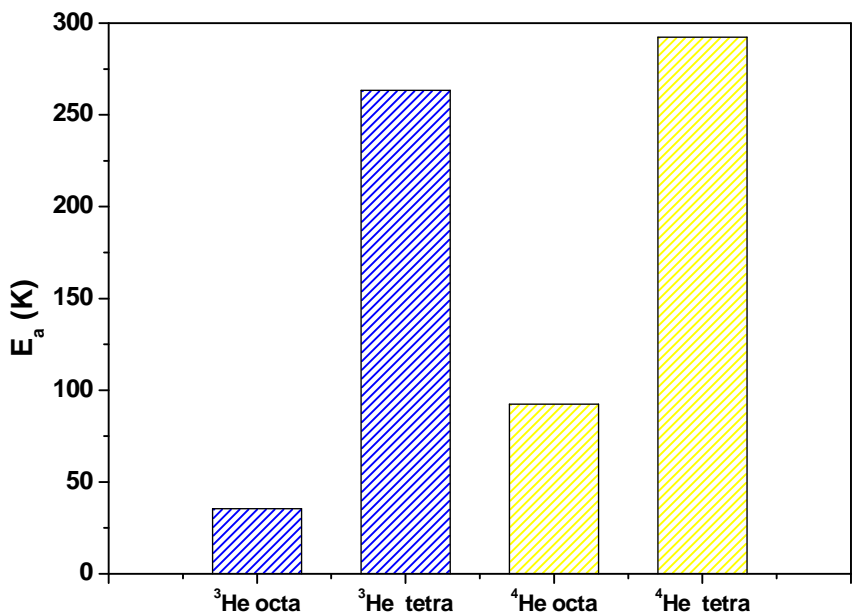

b)

Fig.4. The activation energy of impurity diffusion in fullerite $C_{60}$: a) $^3$He, $^4$He, He$_2$, Ne, Ar in the octahedral subsystem of voids ; b) $^3$He, $^4$He in the octahedral and tetrahedral subsystems of voids.

It is seen in Fig. 4a and Table 1 that the activation energy of impurity diffusion in the octahedral voids of fullerite ($E_{a\,oct}$) increases with the mass and the effective diameter of impurity particles

Table 1. The molecular parameters and activation energies of diffusion of impurity particles in fullerite $C_{60}$.

| Impurity particles | Effective diameter $\sigma$, Å [12] | Mass m, u | $E_{a\,oct}$, K | $E_{a\,tetr}$, K |
|---|---|---|---|---|
| $^3$He | 2.62 | 3 | 35.40 | 263.44 |
| $^4$He | 2.62 | 4 | 92.46 | 292.32 |
| $H_2$ | 2.96 | 2 | 59.94 | - |
| Ne | 2.788 | 20 | 309.43 | - |
| Ar | 3.405 | 40 | 411.17 (> 100 K)<br>46.63 (< 100 K) | - |

The activation energies of impurity diffusion in the tetrahedral subsystem of voids ($E_{a\,tetr}$) were obtained only for helium isotopes. The other impurities demonstrated the saturation kinetics typical for occupation of the octahedral subsystem only [8, 13 – 15].

In terms of energy, the tetrahedral voids are less advantageous for occupation by He atoms [5, 16]. As a result, the values of $E_{a\,tetr}$ exceed considerable those of $E_{a\,oct}$ (Fig. 4).

It is interested that for the octahedral subsystem of voids in $C_{60}$ the activation energy is appreciable lower for $^3$He in comparison with $^4$He (Fig. 4b). In part this may be attributed to the smaller mass of $^3$He atoms and hence the higher energy of zero vibrations of the $^3$He particles in the lattice voids, which suppresses the activation energy. On the other hand, there is a higher probability of quantum effects influencing the diffusion processes for light $^3$He atoms, especially in the octahedral subsystem of voids in which the potential barriers hampering the tunneling of He atoms are lower than in tetrahedral subsystem. These factors become less significant as the barriers hampering the motions of the impurity atoms (molecules) grow higher. In the case of the tetrahedral subsystem with higher potential barriers the difference the activation energies of the $^3$He and $^4$He is much smaller.

## Conclusions

The temperature dependence of the diffusion coefficients of the Ar impurity in the fullerite has been obtained using the measured characteristic times of occupation of the octahedral voids in fullerite $C_{60}$ by Ar atoms. Unlike the He isotope, $H_2$ and Ne impurities, the diffusion coefficients of Ar decrease with lowering temperature in whole interval of investigated temperatures (58-290 K), which corresponds to the thermally activated diffusion of Ar atoms in $C_{60}$. Most likely this behavior of the diffusion coefficients is attributable to the larger mass and effective diameters of Ar atoms comparison with $H_2$, He, and Ne, which reduces the influence of quantum effects on kinetics of $C_{60}$ saturation with argon.

It has been shown that the glass transition in $C_{60}$ decrease sharply (by an order of magnitude) the activation energy of Ar atoms in $C_{60}$, which is most likely connected with the appearance (with the course of the glass transition) of new directions having significantly lower barriers between the interstitial voids in the $C_{60}$ lattice.

We wish to thank Prof. A.I. Prokhvatilov and Prof. Yu. A. Freiman for valuable discussion.

The authors are indebted to the National Academy of Sciences of Ukraine for the financial support of this study in the framework of the special program «Fundamental problems of nanostructural systems, nanomaterials, nanotechnologies».